\documentclass[onecolumn,fleqn,showpacs,showkeys,10pt,superscriptaddress,final]{article}

\usepackage{graphicx}
\usepackage{dcolumn}
\usepackage{bm}
\usepackage{amsmath}
\usepackage{longtable}

\begin{document}

\centerline{\bf  Recent advances in the calculation of dynamical correlation functions}
\vskip 0.5cm

\centerline{J. Florencio$^{1,*}$, O. F. de Alcantara Bonfim$^{2,\dag}$}

\vskip 0.2cm
\centerline{ ${}^1$Instituto de F\'\i sica, Universidade Federal Fluminense, Niter\'oi, Rio de Janeiro, Brazil}

\vskip 0.2cm

\centerline{ ${}^2$Department of Physics, University of Portland, Portland, Oregon, USA }

\vskip 0.2cm

\centerline{ Corresponding authors: $^*$jfj@if.uff.br, $^\dag$bonfim@up.edu} 

\vskip 0.5cm

\centerline{\bf Abstract}
\vskip 0.2cm
\noindent We review various theoretical methods that have been used in recent years to calculate 
dynamical correlation functions of many-body systems.  Time-dependent correlation functions and 
their associated frequency spectral densities are the quantities of interest, for they play a central 
role in both the theoretical and experimental understanding of dynamic properties.  In particular, 
dynamic correlation functions appear in the fluctuation-dissipation theorem, where the response 
of a many-body system to an external perturbation is  given in terms of the relaxation function 
of the unperturbed system, provided the disturbance is small.  The calculation of the relaxation 
function is rather difficult in most cases of interest,  
except for a few examples where exact analytic expressions are allowed. For most of systems
of interest approximation schemes must be used.
The method of recurrence relation has, at its foundation, the solution of Heisenberg
equation of motion of an operator in a many-body interacting system.
Insights have been gained from theorems that were discovered
with that method. For instance, the absence of pure exponential behavior for the
relaxation functions of any Hamiltonian system.
The method of recurrence relations was used in quantum systems such as
dense electron gas,  transverse Ising model,
Heisenberg model, XY model, Heisenberg model with Dzyaloshinskii-Moriya interactions,
as well as classical harmonic oscillator chains. Effects of disorder were
considered in some of those systems.
In the cases where analytical solutions were not feasible, approximation
schemes were used, but are highly model-dependent.  Another important approach
is the numericallly exact diagonalizaton method.  It is used in 
finite-sized systems, which sometimes provides very reliable information of the dynamics 
at the infinite-size limit.  
In this work, we discuss the most relevant applications of the method of recurrence     
relations and numerical calculations based on exact diagonalizations.  
The method of recurrence relations relies on the solution to the coefficients of a continued 
fraction for the Laplace transformed relaxation function.  The calculation of those coefficients 
becomes very involved and, only a few cases offer exact solution.  
We shall concentrate our efforts on the cases where extrapolation schemes must be 
used to obtain solutions for long times (or low frequency) regimes.  We also cover numerical 
work based on the exact diagonalization of finite sized systems.  The numerical work provides 
some thermodynamically exact results and identifies some difficulties intrinsic to the method of 
recurrence relations.

\vskip 1.0cm
\section{Introduction}

\noindent Dynamical correlation functions are central to the understanding of 
time-dependent  properties of many-body systems.  They appear ubiquitously in the
formulation of the fluctuation-dissipation theory, where the response of a system to a weak 
external perturbation is cast in terms of a time-dependent relaxation function 
of the unperturbed system~\cite{Cal51,Kub66}.  

In this article, we are concerned with the recent calculations
of such correlation functions.  We shall cover two lines of approach,
namely the method of recurrence relations and the method of exact diagonalization.

The method of recurrence relations was developed
in the early 1980s~\cite{Lee82a,Lee82b,Lee77,Gri83,Gio83} following the ideas of the 
Mori-Zwanzig projection operator formalism~\cite{Mor65,Zwa61}.  Essentially one solves the
Heisenberg equation of motion for an operator of an interacting system,
from which one obtains dynamic correlation functions, a generalized Langevin
equation, memory functions, etc. 
Review articles found in the literature cover the earlier 
developments~\cite{Lee87,Vis94,Bal03}.

On the other hand, exact diagonalization methods have also
been used in several areas of physics~\cite{Sur75a,Sur75b,Fab97,Boe00a,Gui15}.  
In this method one numerically determines 
the eigenvalues and eigenfunctions of a given Hamiltonian of a finite
system to find the dynamical correlations of interest.
The main drawback is that one is bound by computer
limitations and must deal with finite systems. In addition,
being a numerical method, it does not provide
any new general insight in the form of theorems, etc.
Nevertheless, one can obtain surprisingly good
results which can be readily extended to the
thermodynamic limit.  In a way, exact diagonalization
complements the method of recurrence relations, especially 
when solutions become hard to obtain
by analytic means.  Other approaches can be found in
 Refs.~\cite{Pir88,Pla83a,Bra86,Boh92,Boh93,Bon92,Sto92}.

\section{Dynamical correlation functions}

\noindent Consider a system of $N$ elements such as particles, spins, etc.,
governed by a time-independent Hamiltonian $H$, in thermal equilibrium 
with a heat bath at temperature $T$ .  For two dynamical variables
$X$ and $Y$ of the system, the time-dependent correlation function
is given by the average:
\begin{equation}
<Y(0)X(t)> \, \equiv (1/Z) \, {\rm Tr}\, [ Y(0) X(t) \exp(-\beta H)],
\end{equation}  
\noindent where $\rm Tr [ \dots ]$ denotes a trace over a complete set of states.
Here,  $\beta = 1/k_BT$ is the inverse temperature, 
$Z\, \equiv {\rm Tr}\,\exp(-\beta H)$ is the canonical partition
function, and $X(t)$ is a time-dependent operator 
in Heisenberg representation $X(t) = \exp (iHt/\hbar) X \exp
 (-iHt/\hbar)$,
which satisfies:
\begin{equation}
i \hbar\frac{dX(t)}{dt} = [X(t),H], \qquad X(0) = X,
\end{equation}
where $[X(t),H]$ is the quantum commutator.

In a classical system, the operators are replaced by classical dynamic
variables, the trace by integral over the phase space, and the
commutators by Poisson brackets.

\noindent For a given variable, the time-dependent correlation
function $C(t)$ reads:
\begin{equation}
C(t) = \frac{<X(0)X(t)>}{<X(0)X(0)>}.
\label{eq:C(t)}
\end{equation}
Its Fourier transform $S(\omega)$ is called the spectral density,
or frequency spectrum:
\begin{equation}
S(\omega) = \int_{-\infty}^{\infty} C(t) \exp(-i\omega t)\, dt.
\end{equation}
If we use the integral representation of the Dirac $\delta$-function:
\begin{equation}
\delta(t) = \frac{1}{2 \pi} \int_{-\infty}^{\infty} \exp(- i\omega t)\, d\omega,
\end{equation}
then we obtain
\begin{equation}
C(t) = \frac{1}{2 \pi} \int_{-\infty}^{\infty} S(\omega) \exp(i \omega t) \, d\omega.
\end{equation}
Since the Hamiltonian is time-independent, it follows that $C(t)$ in Eq.~(\ref{eq:C(t)}),
has the property $<X(0)X(t)> \, = \, <~X(\tau) X(t + \tau)>$. If we take $\tau = -t$,
then $ <~X(0)X(t)>\,= \, <X(-t)X(0)>$. Also, it  follows that $S(\omega)$ is real.
Due to the invariance of the trace under cyclic permutations, one finds that
$S(-\omega) = \exp(-\beta \hbar \omega) S(\omega)$.  In the classical limit
($\hbar= 0$ ) or at infinite temperature $\beta = 1/k_BT = 0$, it follows that
 $S(\omega)$ is even in $\omega$.  In general,
the asymmetry in $S(\omega)$ is a typical quantum feature, and is 
referred to as the detailed balance.  

Dynamical correlation functions appear in the 
{\em relaxation function} $R(t)$
from linear response theory~\cite{Kub57,Kub66}:
$$
R(t) = \int_0^{\beta} d\lambda  {<} \exp(\lambda H) Y \exp(-\lambda H)X(t) {>}
$$
\begin{equation}
  \qquad \qquad \qquad \qquad  - \beta <X><Y> 
\end{equation}
where $<\dots>$ is a canonical average, and $X$ and $Y$ are operators.

Time-dependent correlation functions appear in the dynamical structure
factor, are related to the inelastic neutron-scattering cross section,
where the neutron energy changes upon the scattering process.
For a system of interacting spins on a lattice, the dynamic structure 
factor reads:
\begin{equation}
S^{\alpha}(q,\omega) = \sum_n \int_{-\infty}^{\infty} dt \exp[i(qn-\omega t)]<S_j^{\alpha}(0)S_{j+n}^{\alpha}(t)>.
\end{equation}
where $S$ are spin variables and the sum runs over all the lattice sites.

In light scattering experiments, the scattered intensity is given by the
differential cross section, proportional to:
\begin{equation}
I({\mathbf k}, \omega) = \int_{-\infty}^{\infty} dt \exp(-i\omega t)<A_k^{\dagger}(0)A_k(t)>.
\end{equation}
where the form of operator $A$ is system dependent. It also depends on the the particular frequency
of the incoming light.

\subsection{The method of recurrence relations}

The time evolution of a Hermitian operator $A(t)$
is governed by the Heisenberg equation:
\begin{equation}
\frac{dA(t)}{dt} = i {\cal L} A(t),
\label{eq:Heisenberg}
\end{equation}
where:
\begin{equation}
{\cal L} A(t) \equiv HA(t) - A(t)H = [H, A(t)].
\label{eq:calL}
\end{equation}
Consider a time-independent and Hermitian Hamiltonian $H$. 
From now on we will be using a system of units in which $\hbar = 1$.
We seek a solution to Eq.~(\ref{eq:Heisenberg}) for $t \ge 0$, 
thus we set $A(t)=0$ for $t < 0$.

In the method of recurrence relations, the formal solution:
\begin{equation}
A(t) = \exp(iHt)A\exp(-iHt)
\end{equation}
is cast as an orthogonal expansion in a realized Hilbert space $\cal S$
of $d$ dimensions.
That Hilbert space $\cal S$ is realized by the scalar product:
\begin{equation}
(X, Y) = \beta^{-1} \int_0^{\beta} d\lambda \, <X(\lambda)Y> - <X><Y>.
\end{equation}
where $X$, $Y$ $\subset \cal S$, $\beta$ is the inverse temperature, 
$X(\lambda) = \exp(\lambda H) X \exp(-\lambda H)$, and $< \dots>$ denotes
 a canonical ensemble average.

Thus, the time evolution of $A(t)$ is written as:
\begin{equation}
A(t) = \sum_{\nu = 0}^{d-1} a_{\nu}(t) f_{\nu}.
\label{eq:A(t)}
\end{equation}
where $\{ f_{\nu}\}$ is a complete set of states in $\cal S$, while the time-dependence
is carried out by the coefficients $a_{\nu}(t)$.  The dimensionality $d$ of the realized
Hilbert space $\cal S$ is still unknown, but it will be determined later.
If $d$ turns out to be finite, the solutions are oscillatory functions. However,
in most interesting cases $d$ is infinite.  The method of recurrence relations
imposes constraints on which type of solutions are admissible.

By choosing the basal vector $f_0 = A(0) = A$, the remaining basis vectors
are obtained following the Gram-Schmidt orthogonalization procedure, which
is equivalent to the recurrence relation:
\begin{equation}
f_{\nu + 1} = i{\cal L}f_{\nu} + \Delta_{\nu} f_{\nu - 1}, \qquad \nu \ge 0
\label{eq:RRI}
\end{equation}
with $f_1 \equiv 0$, $\Delta_0 \equiv 0$.  The quantity $\Delta_{\nu}$
is defined as the ratio between the norms of consecutive basis vectors:
\begin{equation}
\Delta_{\nu} = \frac{(f_{\nu},f_{\nu})}{(f_{\nu - 1},f_{\nu-1})} \qquad \nu \ge 1.
\end{equation}
The $\Delta$'{}s are referred to as the {\em recurrants} whereas
 Eq.~(\ref{eq:RRI}) is termed the
{\em first recurrence relation}, or  RRI.
The time-dependent correlation function $C(t)$  is given by:
\begin{equation}
C(t) = \frac{<A(0)A(t)>}{<A(0)A(0)>} = (f_0,A(t)) = a_0(t).
\end{equation}
The basal coefficient $a_0(t)$ is just the time-dependent correlation
function.

The time-dependent coefficients $a_{\nu}(t)$ obey a {\em second recurrence
relation} (RRII):
\begin{equation}
\Delta_{\nu + 1} a_{\nu + 1}(t) = - \dot{a}_{\nu}(t) + {a}_{\nu-1}(t) \qquad \nu \ge 0,
\end{equation}
where $\dot{a}_{\nu}(t) = da_{\nu}(t)/dt$, and $a_{-1} \equiv 0$.
It follows from Eq.~(\ref{eq:A(t)}) that the initial choice $f_0 = A(0)$
implies $a_{0}=1$ and $a_{\nu}(0) = 0$ for $\nu \ge 1$. 
Thus the complete time evolution of $A(t)$ is obtained by 
the two recurrence relations RRI and RRII.  One should note that only in 
very few cases a closed analytic solution to a model can be found. More often,
as in many-body problems, approximations are required.

A generalized Langevin equation can be derived for $A(t)$~\cite{Lee82a,Lee82b,Lee83a}:
\begin{equation}
\frac{dA(t)}{dt} = \int_0^t dt'\,\phi(t-t')A(t') = F[t]
\label{eq:Langevin}
\end{equation}
where $\phi$ is the {\em memory function} and $F[t]$ the {\em random force}.
The random force is given as an expansion in the subspace of $\cal S$:
\begin{equation}
F[t] = \sum_{\nu=1}^{d-1} b_{\nu}(t) f_{\nu},
\label{eq:F[t]}
\end{equation}
where the coefficients $b_{\nu}$ satisfy the convolution equation:
\begin{equation}
a_{\nu}(t) = \int_0^t dt' b_{\nu}(t - t') a_0(t'), \qquad \nu \ge 1.
\end{equation}
The memory function $\phi(t)$ is $\phi(t) = \Delta_1 b_1(t)$.
The remaining $b_{\nu}$'s, $b_2$, $b_3$,  are the second
memory function, the third memory function, $\dots$,  etc. 

Consider now the Lapace transform $a_{\nu}(z)$ of $a_{\nu}(t)$:
\begin{equation}
a_{\nu}(z) = \int_0^{\infty} dt \exp(-zt) a_{\nu}(t), \quad {\rm Re}\, z > 0.
\end{equation}
Then RRII can be transformed in the following way:
\begin{eqnarray}
1 &&= a_0(z) + \Delta_1 a_1(z), \\
a_{\nu - 1}(z) &&= a_{\nu}(z) + \Delta_{\nu+1} a_{\nu+1}(z), \quad  \nu \ge 1.
\end{eqnarray}
These equations can be solved for $a_0(z)$:
\begin{equation}
a_0(z) = 1/(z+ \Delta_1/(z + \Delta_2/z + \dots \Delta_{d-1}/z)),
\label{eq:a_0(z)}
\end{equation}
resulting in a continued fraction.  As can be seen from Eq.~(\ref{eq:A(t)}) and the recurrence
relation RRII, that the time-dependence actually depends on the recurrants $\Delta_{\nu}$ only.  
Therefore, the knowledge of all recurrants provides the necessary means to obtain
the time correlation function.  Moreover, the structure
of RRII must be obeyed by time correlation functions. Thus,
a pure exponential decay as well as special polynomials can be
ruled out as solutions, since their recursion relations are not
congruent to RRII.  Also, from  RRII
one obtains $(da_0(t)/dt)|_0 = 0$, which precludes a pure time exponential
as well as other functions that do not have zero derivative at $t=0$.
The method of recurrence relations have since been applied to a variety
of problems, such as the electron gas~\cite{Lee84a,Lee84b,Lee85,Sha92},
harmonic oscillator 
chains~\cite{Flo85,Wie10,Wie12,Yu14, Yu15,Yu16a,Yu16b,Lee16,Yu17,Yu18},
many-particle systems~\cite{Saw02,Sen06,Mok13,Mok15},
spin chains~\cite{Flo87,Flo88,Saw99,Flo00,Nun03,Nun04,Pla83b,Vis90,Mul81,Saw01,Saw03,Sen93,Liu06,Yua10,Li13,Nun20}, 
plasmonic Dirac systems~\cite{Sil15a,Sil15b}, etc.

\subsection{The method of exact diagonalization}

\noindent Given a system governed by a Hamiltonian $H$, one wishes
to numerically determine  the time correlation function $C(t)$, defined by:
\begin{equation}
C(t) = \frac{<A(0)A(t)>}{<AA>},
\label{eq:C(t)b}
\end{equation}
 where $A(t) = \exp(iHt) A \exp(-iHt) $, $\hbar=1$, and the brackets denote
 canonical averages.  We consider here self-adjoint operators
 $A$ and  the Hamiltonian $H$.
 One numerically diagonalizes $H$ and then uses  its eigenvalues  $E_n$ and 
 eigenvectors $|n>$,  $H|n> = E_n|n>$,   to calculate $C(t)$ in Eq.~(\ref{eq:C(t)b}),
where:
\begin{equation}
<AA> = \frac{1}{Z}\sum_n \exp(-\beta H) <n|A^2|n>,
\end{equation}
\begin{equation}
<A(0)A(t)> = \frac{1}{Z} \sum_{n,m} e^{-\beta E_n} e^{- i (E_n - E_m)} |<n|A|m>|^2,       
\end{equation}
with the partition function $Z = \sum_n \exp(-\beta E_n$).  Notice that
the time correlation function is normalized to unity at $t=0$, that is, $C(0) = 1$.

Another quantity of interest is the moment $\mu_k$, also referred 
to as the {\em frequency moments}
which can be obtained from the Taylor
expansion of $C(t)$ about $t=0$:
\begin{equation}
C(t) = \sum_{k=0}^{\infty} \frac{(-1)^k} { (2k)!} \mu_{2k} t^{2k}.
\end{equation}
Since $C(0)=1$ , it follows that $\mu_0 = 1$.  
The moments are given by:
\begin{equation}
\mu_{2k} = \frac{1}{Z} {\rm Tr}\, [e^{-\beta H} A {\cal L}^{2k} A],
\end{equation}
where $\cal L$ is the Liouville operator, Eq.~(\ref{eq:calL}).

From the moments, one can use conversion formulas to obtain
the recurrants $\Delta$'s of the method of recurrence relations
from the frequency moments~\cite{Vis94}.
Suppose the moments $\mu_0=1$ and $\mu_{2k}$, $k= 1, \dots, K$ are known.
The first $K$ recurrants $\Delta_{\nu}$ are determined by the equations:
\begin{equation}
\Delta_{\nu} = \mu_{2\nu}^{(\nu)}, \qquad \mu_{2k}^{(\nu)} = \frac{ \mu_{2k}^{(\nu - 1)}}{\Delta_{\nu - 1}} - \frac{\mu_{2k-2}^{(\nu - 2)}}{\Delta_{\nu - 2}}, \quad ,
\label{eq:Delta}
\end{equation}
for $k = \nu, \nu + 1, \dots, K$ and $\nu = 1, 2, \dots, K$, with $\mu_{2k}^{(0)}= \mu_{2k}$, $\Delta_{-1} = \Delta_0 = 1$,
$\mu_{2k}^{(-1)} = 0.$s

For instance, if the first moments $\mu_0 =1$, $\mu_2$, $\mu_4$, $\dots$, $\mu_{10}$ are given, the
recurrences are obtained from Eq.~(\ref{eq:Delta}):
\begin{eqnarray}
&&\Delta_1 = \mu_2, \nonumber \\
&&\Delta_2 = -\mu_2 + \mu_4/\mu_2, \nonumber \\
&&\Delta_3 = \mu_4 (\mu_4/\mu_2 - \mu_6/\mu_4)/\mu_2 (\mu_2 - \mu_4/\mu_2), \nonumber \\
&&\Delta_4 = - \mu_4/\mu_2 + \mu_6/\mu_4 \nonumber \\
&&- \mu_4 (\mu_4/\mu_2 - \mu_6/\mu_4)/ \mu_2 (\mu_2 - \mu_4/\mu_2) \nonumber \\ 
&&+ \mu_6 (\mu_6/\mu_4 - \mu_8/\mu_6)/ \mu_4 (\mu_4/\mu_2 - \mu_6/\mu_4).
\end{eqnarray}

Conversely, suppose one has the first $K$ known recurrants, $\Delta_{\nu}$, $\nu = 1, 2, \dots, K$, and
$\Delta_{-1} = \Delta_0 = 1$. Then, the  moments $\mu_{2 \nu}^{(0)} = \mu_{2 \nu}$ are obtained from the following conversion formula:
\begin{equation}
\mu_{2k}^{(\nu - 1)} = \Delta_{\nu - 1} \mu_{2k}^{(\nu)} + \frac{\Delta_{\nu-1}}{\Delta_{\nu - 2}} \mu_{2k - 2}^{(\nu - 2)},
\label{eq:Moment}
\end{equation}
for $\nu = k, k-1, \dots, 1$ and $k = 1, 2, \dots, K$, with $\mu_{2k}^{(-1)}~=~0$.

In case the first recurrants $\Delta_1$, $\Delta_2$, $\dots$, $\Delta_4$, are known,  
the moments $\mu$ are found to be:
\begin{eqnarray}
&&\mu_0 = 1,  \nonumber \\
&&\mu_2 = \Delta_1,   \nonumber \\
&&\mu_4 = \Delta_1(\Delta_1 + \Delta_2),  \nonumber \\
&&\mu_6 =  \Delta_1((\Delta_1 + \Delta_2)^2 + \Delta_1 \Delta_3), \nonumber \\
&&\mu_8 = \Delta_1 \left( (\Delta_1+\Delta_2)^2 + \Delta_2 \Delta_3 \right)  \nonumber \\
&& \times \left( \Delta_1 + \Delta_2 + \frac{\Delta_2 \Delta_3}{\Delta_1 + \Delta_2} 
    + \Delta_2 \Delta_3 \frac{\Delta_3 + \Delta_4 + \frac{\Delta_2 \Delta_3}{\Delta_1 + \Delta_2}}{(\Delta_1 + \Delta_2)^2 + \Delta_2 \Delta_3} \right). \nonumber \\
&&{ }
\end{eqnarray}

Typically, the analytical determination of the recurrants becomes increasingly time consuming.
In practice, only a few of them can be obtained to be used in  an extrapolation scheme
 to obtain  higher-order recurrants.  Several extrapolation schemes have been used.
One of the  simplest  is to set the unknown recurrants
to zero, thus truncating the continued fraction for $a_0(z)$,
which leads to a finite number of poles in the complex plane~\cite{Pir88}.
In other problems, it is most appropriate to introduce a Gaussian termination,
that is, a sequence of recurrants $\Delta{\nu}$ that grow linearly with its index $\nu$,
$\Delta_{\nu} = \nu \Delta$~\cite{Flo87,Vis94}.  
Other extrapolation schemes are tailored to the problem at hand, especially
if the recurrants are not expected to grow  indefinitely.

\section{Applications to interacting systems}

\noindent The dynamics of spin chains has attracted a great deal of attention
in recent decades. Exact results for the longitudinal dynamics of the
one dimensional XY model have been obtained with the Jordan-Wigner
transformation~\cite{Nie67}. Later, exact results for the transverse time correlation 
functions of the XY and the transverse Ising chain were obtained at the high 
temperature limit by using different methods~\cite{Bra76,Cap77,Flo87}.

A great deal of progress was achieved in the calculations
of the dynamic correlation functions of spin models in one dimension.
It was soon recognized that exact solutions using the method of
recurrence relations were difficult to obtain, however a notable exception 
is the classical harmonic oscillator chain where the time correlation functions 
were obtained exactly~\cite{Flo85}.

The problem of a mass impurity in the harmonic chain was solved later,
and its dynamical correlation functions were found to have the same form 
as in the quantum electron
gas in two dimensions, thus showing that unrelated quantities in these two 
models displayed the
same dynamical behavior, that is, the have {\em dynamic equivalence}~\cite{Lee89}.
It should be mentioned that harmonic oscillator chains have been the subject of
a considerable amount of work with the method of recurrence 
relations~\cite{Wie10,Wie12,Yu14, Yu15,Yu16a,Yu16b,Lee16,Yu17,Yu18}.

The method of recurrence relations provides
 important insights on how to proceed to obtain reliable approximate
solutions.  The cornerstone quantity in the dynamics is the recurrant,
which is the only quantity that ultimately determines the dynamics of the
model.  Often it is only possible to determine a few of the recurrants analytically. 
The calculations become too lenghty so that one must
stop at a given order. Thus,  an extrapolation method must be devised
for the higher order recurrants, which hopefully will
have the essential ingredients to produce reliable time-dependent correlation functions 
for longer times as well as spectral densities with the expected behavior near
the origin 
$\omega =0$~\cite{Sen93,Sen99,Flo92,Sen06}.

The dynamics of the transverse Ising model in two dimensions
was studied with the method of recurrence relations~\cite{Sen93a,Sen93b,Flo95,Che10}.
The dynamic structure factor of that model compares well to the experimental data
of the compound LiTbF${}_4$~\cite{Kot88}. 

The dynamics of spin ladders has also attracted  interest
from researchers.  The dynamical correlation functions were obtained for a two-leg spin
ladder with XY interaction along each leg and interchain Ising couplings in 
a random magnetic field. More recently, the dynamics of a ladder with
Ising couplings in the legs and steps as well as four-spin plaquette interactions
in a magnetic field~\cite{Sou20} have been also investigated.

The dynamical correlations of the Heisenberg model in one dimension
have been have been a subject of great interest in the 
recent decades~\cite{Sur75a,Sur75b,Fab97,Boh92,Kha16}.
The method of recurrence relations has been employed in various 
works~\cite{Lee83,Vis91,Sen02,Sen92,Boh94,Sen95,Liu90}.  In spite of the
progress made thus far, the long-time dynamics of the Heisenberg
spin model is still an open problem.  For instance, there is the standing
problem on the power law exponent $\alpha > 0$ (~$\sim t^{-\alpha}$~) of the time correlation 
function as $t \to \infty$.  From the work of  Fabricius et al.~\cite{Fab97}, 
we find that the time correlation functions of the Heisenberg model decay more slowly
than that in the XY model, for which the exact solution is known, $C(t) \sim t^{-1}$ for large $t$.   
Thus we  infer that the numerical evidence suggests that $\alpha \ge 1$ for
the Heisenberg model.

There has been a great deal of work that uses exact diagonalization
to study the dynamics of spin systems~\cite{Sur75a,Sur75b}.   Earlier works  with  the
Heisenberg model used this technique.
Later on, other systems were scrutinized by using exact diagonalization. 
One of those systems is the Ising model with four-spin interactions in a transverse field. 
The time correlation function was obtained for one dimensional and
infinite temperature~\cite{Flo97}, where the Gaussian behavior shown in the usual
transverse Ising model was ruled out. The effects of disorder on the
dynamics of that model were obtained for the cases where the
random variables are drawn from  bimodal distributions
of random couplings and fields~\cite{Boe00a,Boe00b,Bon01}.
Dynamical correlation functions were also obtained for the
system at finite temperatures, ranging from
$T=0$ to $T= \infty$~\cite{Flo04}.

\subsection{Heisenberg model with Dzyaloshinskii-Moriya interactions}

\noindent The dynamical structure factor for a quantum spin Heisenberg chain with 
Dzyaloshinskii-Moriya (DM) interactions~\cite{Dzy58,Mor60} has been investigated 
 by different approaches, such as spin wave theory~\cite{Xia88}, mean-field~\cite{Lim08}, and projection 
 operator techniques~\cite{Pir00}.
The dynamics of the related XY model with DM interactions was also studied by employing
Jordan-Wigner fermions~\cite{Der05,Der06,Ver07}.

The dynamical correlation functions of  the spin-1/2 Heisenberg model
with DM interactions in a transverse magnetic field was studied recently with
the method of recurrence relations. 
The model  Hamiltonian for a one-dimensional chain is given by:
\begin{eqnarray}
H &&= - J \sum_i (\sigma_i^x \sigma_{i+1}^x + \sigma_i^y \sigma_{i+1}^y
    + \sigma_i^z \sigma_{i+1}^z)  \qquad \qquad  \nonumber \\
         &&     - D\sum_i (\sigma_i^x \sigma_{i+1}^y  - \sigma_i^y \sigma_{i+1}^x) 
    - \sum_i B_i \sigma_i^x,
\label{eq:Hamiltonian-HDMB}
\end{eqnarray}
where $J$ is the Heisenberg coupling, $D$ is the Dzyaloshinskii-Moriya interaction, and $B_i$ is a 
magnetic field perpendicular to the DM axis. The quantities $\sigma_i^{x,y,z}$  are the
usual Pauli operators. 

The effects of a uniform magnetic field $B_j = B$ on the dynamics  are investigated in the infinite temperature limit~\cite{Nun18}.
 The purpose is to determine the  time correlation function $C(t) = <\sigma_j^z \sigma_j^z(t)>$ and its
 associated spectral density $S(\omega)$.
The first four recurrants are determined analytically and an extrapolation scheme 
is devised to obtain higher order recurrants.   Such scheme must take into account 
what is already known from the solutions of related problems.   

One crucial point is to determine whether or not the extrapolated recurrants grow indefinitely.  
The time correlation function of the longitudinal 
spin component in the $XY$ chain is known exactly at $T= \infty$, 
$C(t) = J_0^2(t) \sim t^{-1}$ asymptotically
for large times, where $J_0$ is the Bessel function of first kind~\cite{Nie67}.
In this case, the recurrants tend to a constant finite value as $\nu \to \infty$. 
There are numerical indications that the time correlation function of the
Heisenberg model decays as a power law~\cite{Fab97}, which
suggests that the extrapolated recurrants grow asymptotically to a finite value.
For the Hamiltonian Eq.~(\ref{eq:Hamiltonian-HDMB})  the extrapolation is the
following power-law: 
\begin{equation}
\Delta_{\nu} = \Delta_{\infty} - \frac{b}{\nu^{\beta}}, \quad \nu \ge n_c
\label{eq:Deltaplus}
\end{equation}
where $n_c$ is the order of the last exactly calculated recurrant.
The limit value $\Delta_{\infty}$ is obtained by extrapolating
the last two recurrants to the origin of $1/\nu$.
The constants $b$ and $\beta$ guarantee smooth behavior
of the recurrants above and below $\nu = n_c$.

Once the recurrants are obtained, the relaxation function and its spectral 
density can be readily obtained.
For the special case without DM interaction, the result shows good agreement 
with the known results for the XY and Heisenberg models.  
The full calculation reveals that the effects of the external field are to produce 
stronger and more rapid oscillations in the relaxation functions, as well as a suppression 
of the central peak in the spectral density. In addition a peak centered at a well defined
frequency appears, which is attributed to an enhancement of the collective mode of spins 
precessing about the external field.  It should be noted that the method of recurrence
relations was also used to  study the dynamics of the XY model with DM interaction~\cite{Li13}.

The effects of disorder in a transverse magnetic field on the dynamical correlation
functions are investigated with the bimodal distribution for $B_i$:
\begin{equation}
\rho(\{B_i\}) = \Pi_i [ (1-p) \delta(B_i - B_A) + p \delta(B_i - B_B)].
\label{eq:bimodal}
\end{equation} 
The method of recurrence relations is then applied to obtain the dynamical
correlation functions for a given realization of disorder~\cite{Nun20}.  Next, the 
average over the random fields is performed by using the 
distribution Eq.~(\ref{eq:bimodal}).  This is accomplished by
defining the scalar product in the Hilbert space $\cal S$, so as
to include an average over the random variables in addition
to the thermal average.  Four recurrants $\Delta_{\nu}$ are
obtained, and an extrapolation is made for the remaining
recurrants. In practice only the first dozen are needed
to attain convergence. 

The time correlation function and its associated spectral density
 were obtained for $D=1$ and $B_A= 0$ and $B_B= 4$ in units of 
 the Heisenberg coupling $J$.
 When the probability $p$ is very small, a strong
 central mode appears, as well as a shoulder in the spectral density.
 As $p$ increases, there is supression of the central mode
 as well as the shoulder. 
 On the other  hand, for large values of the probability $p$, a nonzero 
 frequency peak appears, resulting from the precession
 of the spins around the magnetic field adding further suppression of the central peak. 
 This central mode behavior versus collective dynamics, is a known feature 
 of the dynamics of spin systems and they are in some sense universal.
 However, in the present case the appearence of a shoulder for small $p$  is an 
 interesting novel feature.

\subsection{Random transverse Ising model}

\noindent Consider the $s=1/2$ spin model in one dimension:
\begin{equation}
H = - \frac{1}{2}\sum_i J_i \sigma_i^x \sigma_{i+1}^x - \frac{1}{2} \sum_i B_i \sigma_i^z,
\label{eq:RTIM}
\end{equation}
where $J_i$ and $B_i$ are exchange couplings and transverse
fields, respectively.  These couplings and fields are 
random variables drawn from distribution functions. The quantities
$\sigma_i^{\alpha} (\alpha = x,y,z)$ are Pauli matrices.
The model is referred to as the random transverse Ising model (RTIM),
and its dynamical correlation function in the infinite temperature limit 
has been investigated  by using the 
method of recurrence relations~\cite{Flo99}.

The time correlation $C(t)$ is defined by:
\begin{equation}
C(t) = \overline{<\sigma_j^x \sigma_j^x(t)>},
\label{eq:C(t)_RTIM}
\end{equation}
where the line indicates that an average over the random variables is performed
after the statistical average $< \dots >$.
The time evolution of $\sigma_j^z(t)$ in a system governed by
the Hamiltonian Eq.~(\ref{eq:RTIM}) is given as an expansion
in a Hilbert space $\cal S$ of $d$ dimensions, where 
$d$ is to be determined later:
\begin{equation}
\sigma_j^x(t) = \sum_{\nu = 0}^{d-1} a_{\nu}(t)f_{\nu},
\end{equation}
where $f_{\nu}$ are orthogonal vectors spanning $\cal S$.
The time dependence is contained in the coefficients  $a_{\nu}(t)$.

The inner product in $\cal S$ in the infinite temperature limit is defined 
in such a way that it encompasses both the thermal average in
a realization of disorder and the average over the random 
variables:
\begin{equation}
(A, B) = \overline{<A B^{\dagger>}} - \overline{<A><B^{\dagger}>},
\label{eq:scalar-RTIM}
\end{equation}
where $A$ and $B$ are vectors in $\cal S$.
This definition of scalar product ensures that the form of the recurrence 
relations in unchanged.

The zeroth basis vector $f_0$ is chosen as the variable of interest,
$f_0 = \sigma_j^x$.  Thus, the zeroth-order coefficient $a_0(t)$ can
be identified with the time-dependent correlation function of interest:
\begin{equation}
a_0(t) = (f_0,f_0) =  \overline{<\sigma_j^x \sigma_j^x(t)>} = C(t).
\label{eq:a_0(t)-RTIM}
\end{equation}
The remaining basis vectors $f_{\nu}$, $\nu = 1, 2, ..., d-1$, are obtained
from the recurrence relation RRI, Eq.~(\ref{eq:RRI}). The first vectors are then:
\begin{eqnarray}
&& \qquad \qquad \qquad \quad  f_0 =  \sigma_j^x , \nonumber \\
&& \qquad \qquad \qquad f_1 =  B_j \sigma_j^y , \nonumber \\
&& f_2 = (\Delta_1 - B_j^2) \sigma_j^x + B_j J_{j-1}\sigma_{j-1}^x \sigma_j^z 
 + B_j J_j \sigma_j^z \sigma_{j+1}^x, \nonumber \\
&& \qquad  f_3 = - B_j (J_{j-1}^2 + J_j^2 + B_j^2 -\Delta_1 - \Delta_2)\sigma_j^y \nonumber \\
&& \quad - 2 B_j J_{j-1} J_j \sigma_{j-1}^x \sigma_j^y \sigma_{j+1}^x + B_{j-1}J_{j-1} \sigma_{j-1}^y \sigma_j^z \nonumber \\
&& \qquad \qquad \quad + B_jB_{j+1} J_j \sigma_j^z \sigma_{j+1}^y, 
\label{eq:f_2-RTIM}
\end{eqnarray}
etc. The vectors $f_4$, $f_5$, $\dots$, $f_9$ were obtained analytically
but not reported because of their length~\cite{Flo99}. 
However, they were used in all of the subsequent calculations.  
The first three recurrants are the following,
\begin{eqnarray}
&&  \qquad \qquad \qquad \qquad  \Delta_1 = \overline{B_j^2}, \nonumber \\ 
&& \qquad \qquad  \Delta_2 = 2 \overline{J_j^2} - \overline{B_j^2} + \overline{B_j^4}/\overline{B_j^2}, \nonumber \\
&& \Delta_3 = \frac{\overline{B_j^6} 
+ 2 \overline{J_j^2}^2 \,  \overline{B_j^2} + 2\overline{J_j^4}\overline{B_j^2}  + 2 \overline{J_j^2}\,\overline{ B_j^2}^2   - \overline{B_j^4}^2/ \overline{B_j^2}}{  2 \overline{J_j^2} \overline{B_j^2} - \overline{B_j^2}^2 + \overline{B_j^4}   }. \nonumber \\
{\ \  } &&{} {\ \ }
\label{eq:Delta-RTIM}
\end{eqnarray}
Notice that the couplings and fields are site-dependent.

There are two types of disorder considered in Ref.~\cite{Flo99}, random fields and
random spin couplings.  Each case is treated separately.
In both cases a simple bimodal distribution is used for the random variable.
 The field $B_i$ (or the coupling $J_i$) can assume two distinct values,
  with probalities $q$ ($p$) and $1-q$ ($1-p)$, respectively. 
 The time correlaton function and the spectral density are then obtained numerically.  
 For the pure cases, ($p=q=1$),  two types of behavior emerge, depending on the 
 relative strength between $J$ and $B$.
For $J>B$, the dynamics is dominated by a central-mode behavior, whereas for $J<B$
a collective-mode is the prevailing dynamics.  In the disordered cases, the dynamics
is neither central-mode nor collective-mode type, but something in between those
types of dynamics.

\subsection{Transverse Ising model with next-to-nearest neighbors interactions}

\noindent Consider the transverse Ising model with an additional 
axial next-nearest-neighbor interaction (transverse ANNNI model)~\cite{Gui15}. 
 The Hamiltonian  for a chain with $L$ spins can be written as:
\begin{equation}
H =  J_1 \sum_{i=1}^{L}\,\sigma_{i}^{z}\sigma_{i+1}^{z}
     - J_{2}\sum_{i=1}^{L}\,\sigma_{i}^{z}\sigma_{i+2}^{z}
     - B\sum_{i=1}^{L}\,\sigma_{i}^{x}~,
\label{eq:Hamiltonian-TINNN}
\end{equation}
where $\sigma_i^{\alpha}$ are the usual spin-1/2 operators, $\alpha = x, y, z$. Periodic boundary conditions 
are imposed on this model, namely $\sigma_{i+L}^{\alpha} = \sigma_{i}^{\alpha}$. 
Consider antiferromagnetic ($J_1 >0)$ Ising interactions. A competing ferromagnetic interaction 
is assumed for the next-nearest-interaction ($J_2 > 0$). The transverse magnetic field ($B$) induces 
the quantum fluctuations. In what follows we set $J_1 = 1$ as the unity of energy. 

In the absence of a transverse magnetic field and of thermal fluctuations ($T=0$) the ground-state 
properties of the model are exactly soluble and several phases are present.  For $J_2 < 0.5$ the ground state is ordered ferromagnetically.  For $J_2 > 0.5$, a phase consisting of two up-spins followed by two down-spins is periodically formed. 
The phase is known as $<2,2>$-phase or an anti-phase. For $J_2 = 0.5$, the model has a multiphase point. The ground-state is highly degenerate with many phases of the type $<p,q>$ corresponding to a periodic phase with $p$-up spins followed by $q$-down spins, among other spin
configurations. The number of degenerate states increases exponentially with the size of the system $L$.
In the case where $J_2 = 0$ and the magnetic field is switched on, the model becomes the Ising model
in a transverse field which was exactly solved by Pfeuty~\cite{Pfe70}. In this model, due to quantum fluctuations induced by the transverse magnetic field, a second order phase transition occurs at $B= 1$, which separates a ferromagnetic phase at low magnetic fields from a paramagnetic phase at high magnetic  fields. For the full Hamiltonian, Eq.~({\ref{eq:Hamiltonian-TINNN}), the competing interaction between the ferromagnetic and antiferromagnetic terms induces frustration in the magnetic ordering. 
This will give rise to a much richer variety of phases when either the transverse magnetic field or the spin-spin interations are varied, such as ferromagnetic or antiferromagnetic phases, disordered or paramagnetic phases, and floating phases~\cite{Gui15}.  Such variety of phases in the ground-state could carry over their effects into the dynamics at the high temperature limit, like the known transverse Ising model.  In this model, a signal of the ground-state transition is manifested in the Gaussian behavior at criticality of the dynamical correlation functions at $T=\infty$~\cite{Flo87}. 

The main quantity of interest is the time-dependent correlation function:
\begin{equation}
C(t) = <\sigma_j^x(0) \sigma_j^x(t) >,
\label{eq:C(t)-ANNNI}
\end{equation}
where $\sigma_j^x(t) = e^{iHt} \sigma_j^x e^{-iHt}$ and $< \cal O>$ is a
canonical average of the operator $\cal O$.  The method of exact diagonalization
will be employed  to study the dynamics,  however, the recurrants of the
method of recurrence relations will also be obtained.

The numerical calculations will be performed 
at the high-temperature limit,  $T=\infty$,  hence:
\begin{equation}
C(t)  = \frac{1}{2^L} {\rm Tr}\, (\sigma_j^x e^{iHt} \sigma_j^x e^{-iHt}).
\label{eq:C(t)-infty-ANNNI}
\end{equation}
One of the properties of $C(t)$ is that it is real and an even function of the time $t$. 
Therefore, the  Taylor expansion about $t=0$ has only even powers of $t$:
\begin{equation}
C(t) = \sum_{k=0}^{\infty} \frac{(-1)^k}{(2k)!} \mu_{2k}\, t^{2k},
\label{eq:Taylor}
\end{equation}
where the frequency moments are expressed in terms of
the trace over iterated commutators:
\begin{equation}
 \mu_{2k} = \frac{1}{2^L} {\rm Tr}\, (\sigma_j^x {\cal L}^{2k} \sigma_j^x),
\label{eq:mu}
\end{equation}
with ${\cal L}$ defined such that: 
\begin{equation}
{\cal L}A = [H, A] = HA - AH,
\label{eq:Liouville}  
 \end{equation} 
where $H$ is the Hamiltonian and $A$ an operator.  

\begin{figure}
\includegraphics[width=8.0cm, height= 6.0cm, angle=0]{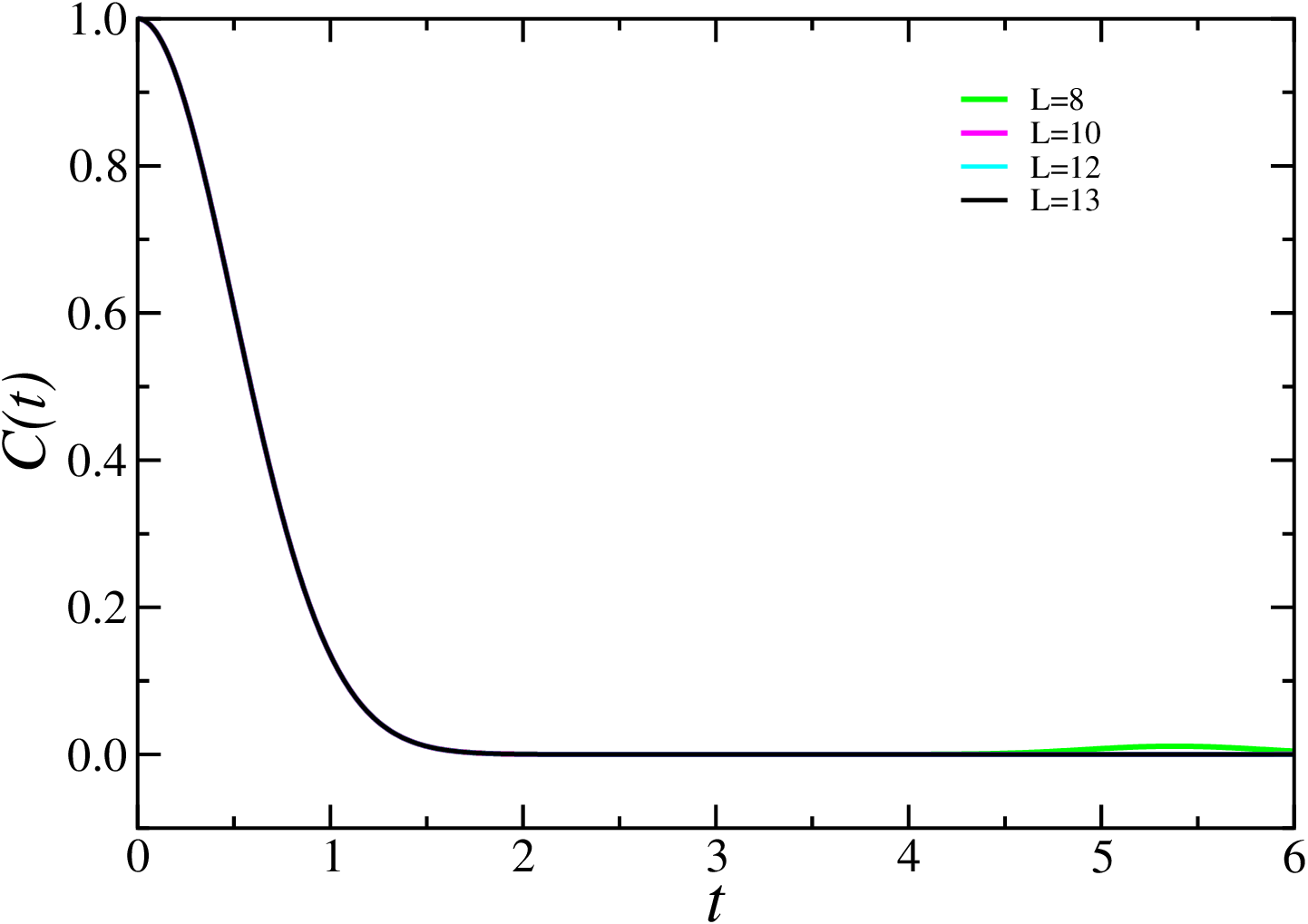}
\caption{
Time correlation function of a tagged spin in the TI model
when $B = J =1.0$ for some chain sizes $L$, as indicated.
Here and in the next figures $J = 1$ is the energy unit.
The exact solution is a Gaussian, which lies underneath
the $L=13$ curve.
\label{fig:fig1}
}
\end{figure}

The correlation function is calculated in the Lehman representation.
First, we consider the energies $E_n$ and eigenstates $|n>$ of the Hamiltonian,
obtained from the eigenvalue equation $H|n> = E_n |n>$. Then, the correlation 
function takes the form:
\begin{eqnarray}
  C(t)&&= \frac{1}{2^L}\sum_{m,n}
               \cos (E_{n}-E_{m})t |<n|\sigma_{j}^{x} |m>|^{2}, \nonumber \\
         &&= \sum_{k=0}^{\infty} \frac{(-1)^k}{(2k)!} \mu_{2k} t^{2k},
\label{eq:Cexplicit}
\end{eqnarray}
where the moments $\mu_{2k}$ are given by:
\begin{equation}
  \mu_{2k} = \frac{1}{2^L}\sum_{m,n}
               (E_{n}-E_{m})^{2k} |  <n|\sigma_{j}^{x} |m>|^{2}.
\label{eq:muexplicit}
\end{equation}

The spectral density $S(\omega)$ is simply the Fourier transform of $C(t)$:
\begin{equation}
S(\omega) = \int_{-\infty}^{\infty}\,C(t)e^{-i\omega t}dt.
\label{eq:S(w)}
\end{equation}
After using  Eq.~(\ref{eq:Cexplicit}), the spectral density can be cast in the form:
\begin{equation}
S(\omega) = \frac{\pi}{2^L} \sum_{m,n}|<n|\sigma_j^x|m>|^2 
[\delta(\omega-\epsilon_{nm}) + \delta(\omega +\epsilon_{nm})],
\label{eq:Sdirac}
\end{equation}
where $\epsilon_{nm} \equiv E_n - E_m$.

The Dirac $\delta$-function is approximated
by  a rectangular window of width $a$ and unit area, centered at the zeros of their arguments.  
The width $a$, can be adjusted to reduce fluctuations.  
Another approach could be the use of histograms, such as in Ref.~\cite{Flo04}.
However,  the general shape of the spectral density
$S(\omega)$ is the same, although the rectangle approximation gives more accurate
results. Therefore, both dynamical correlation functions $C(t)$ and $S(\omega)$
can be calculated directly via exact diagonalization.

\begin{figure}
\includegraphics[width=8.0cm, height= 6.0cm, angle=0]{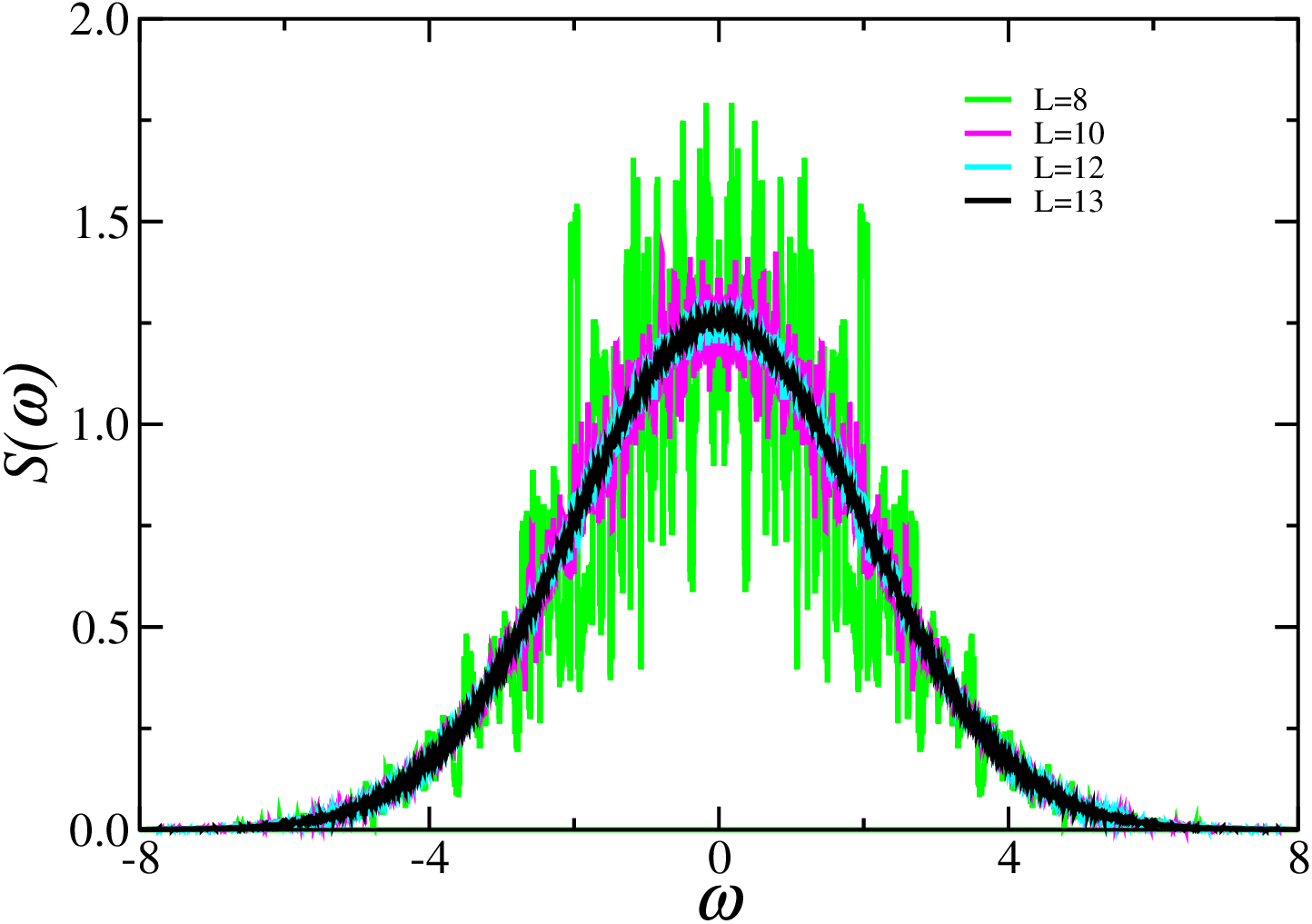}
\caption{Spectral density for the  TI model  ($B = 1$ and $J_2=0$)
and different chain sizes. The plots are the time Fourier transforms of 
the curves in Fig.~\ref{fig:fig1}. The curves for finite chains oscillate
 around the exact Gaussian result of the infinite chain.
\label{fig:fig2}
}
\end{figure}

As a case test, Guimar\~aes et al.~\cite{Gui15} consider $B=1$ and $J_2=0$,
the usual transverse Ising model (TIM) with dynamical correlation
functions known exactly in the high temperature limit~\cite{Cap77, Flo87}.
Figure~\ref{fig:fig1} shows their numerical results for the time correlation function
for $B = 1$ and several lattice sizes. The results for $L=12, 13$,  agree very
well with the exact result of the infinite system, $C(t) = \exp (-2t^2)$ in
the time interval of interest $0 \le t \le 10$.
Convergence toward the thermodynamic result increases as the system size grows.  
However, already for $L=13$ the numerical calculations reproduce the Gaussian behavior
found by the exact calculation.  The corresponding spectral density is
shown in Fig.~\ref{fig:fig2} for different
chain sizes.  The Dirac $\delta$ functions are approximated
by a rectangle of unit area and width $a=0.1$. That is
the best value for the width $a$ to reduce the fluctuations
due to finite-size effects.  Those fluctuations decrease
in amplitude as one considers larger system sizes.
The frequency-dependent Gaussian of the exact result 
is already masked by the curve for $L=13$.  Therefore,
the method works just fine with the transverse Ising
model, and very likely will do so with the transverse
ANNNI model.

In the following, consider the representative cases
$B=0.5$, $1.0$, and $2.0$. These cases should cover
the relevant possibilities for $B$ in the transverse ANNNI model.
Consider first $B=0.5$.
The time correlation function $C(t)$ is shown in Fig.~\ref{fig:fig3}
for different next-to-nearest neighbor couplings $J_2$.
There are pairs of curves for a given $J_2$, dashed lines for $L=12$
and solid lines for $L=13$.   Those two lines agree very well
with each other for the range of time $t$ displayed.
The quantitative agreement between the $L=12$ and $L=13$
curves is an indication that within the accuracy used,
the thermodynamic value has already been obtained.
The features shown in Fig.~\ref{fig:fig3} are real
and will not change in the thermodynamic limit.
They possibly could be traced back to the rich
ground-state phase diagram,  however, a careful
investigation is still necessary to clarify that point.

In general, the decay of $C(t)$ with time is slower
for larger $J_2$.  The corresponding spectral
density is displayed in Fig.~\ref{fig:fig4}, calculated
for $L=13$.  Other than the height near the origin $\omega = 0$,
the remaining plots should not change essentially for
larger chain sizes, or at the thermodynamic limit.
The distinctive feature is the enhancement of
the central mode as $J_2$ increases.  

\begin{figure}
\includegraphics[width=8.0cm, height= 6.0cm, angle=0]{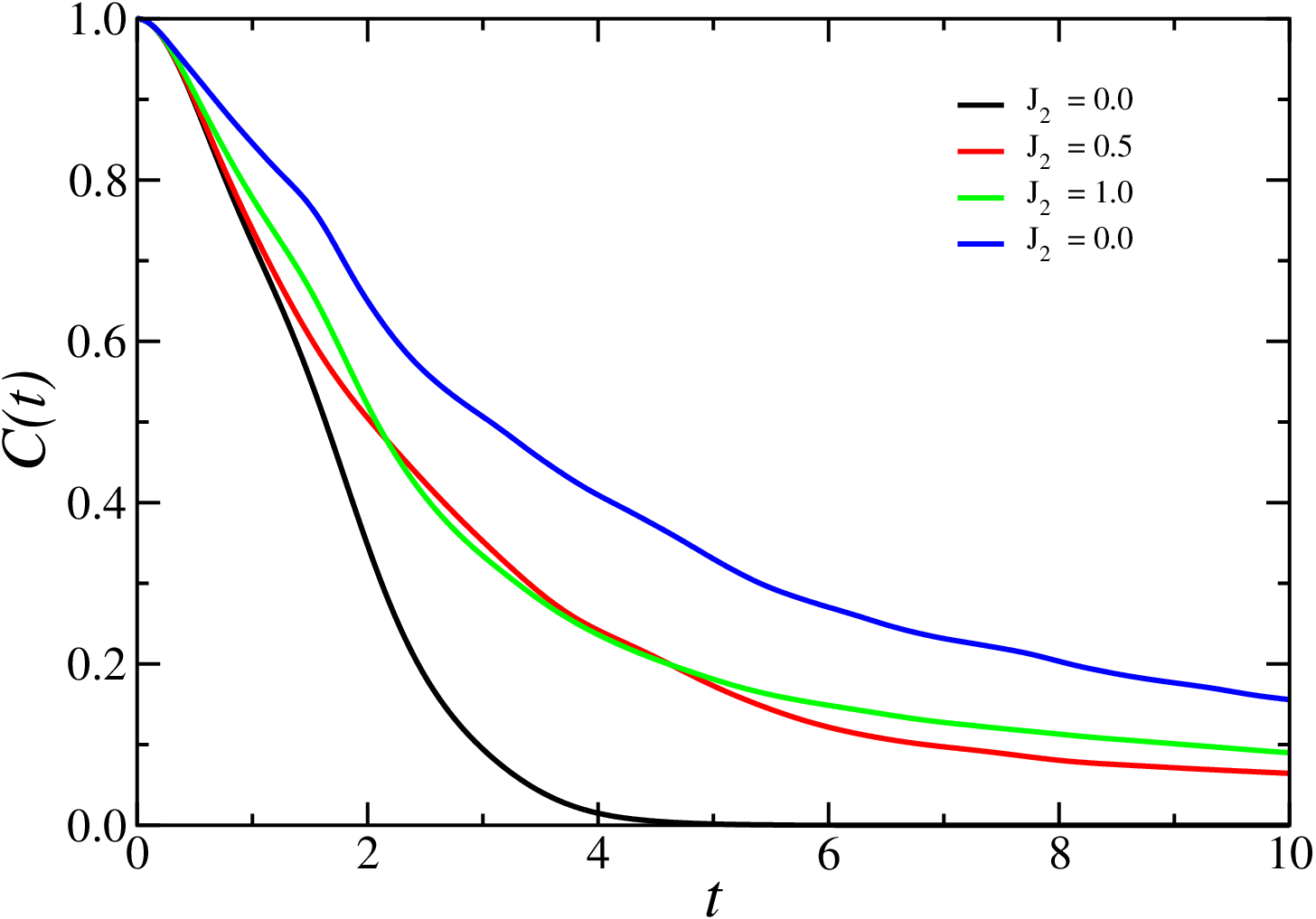}
\caption{Time-dependent correlation function for $B=0.5$ and 
several values of the NNN coupling $J_2$.
\label{fig:fig3}
}
\end{figure}

\begin{figure}
\includegraphics[width=8.0cm, height= 6.0cm, angle=0]{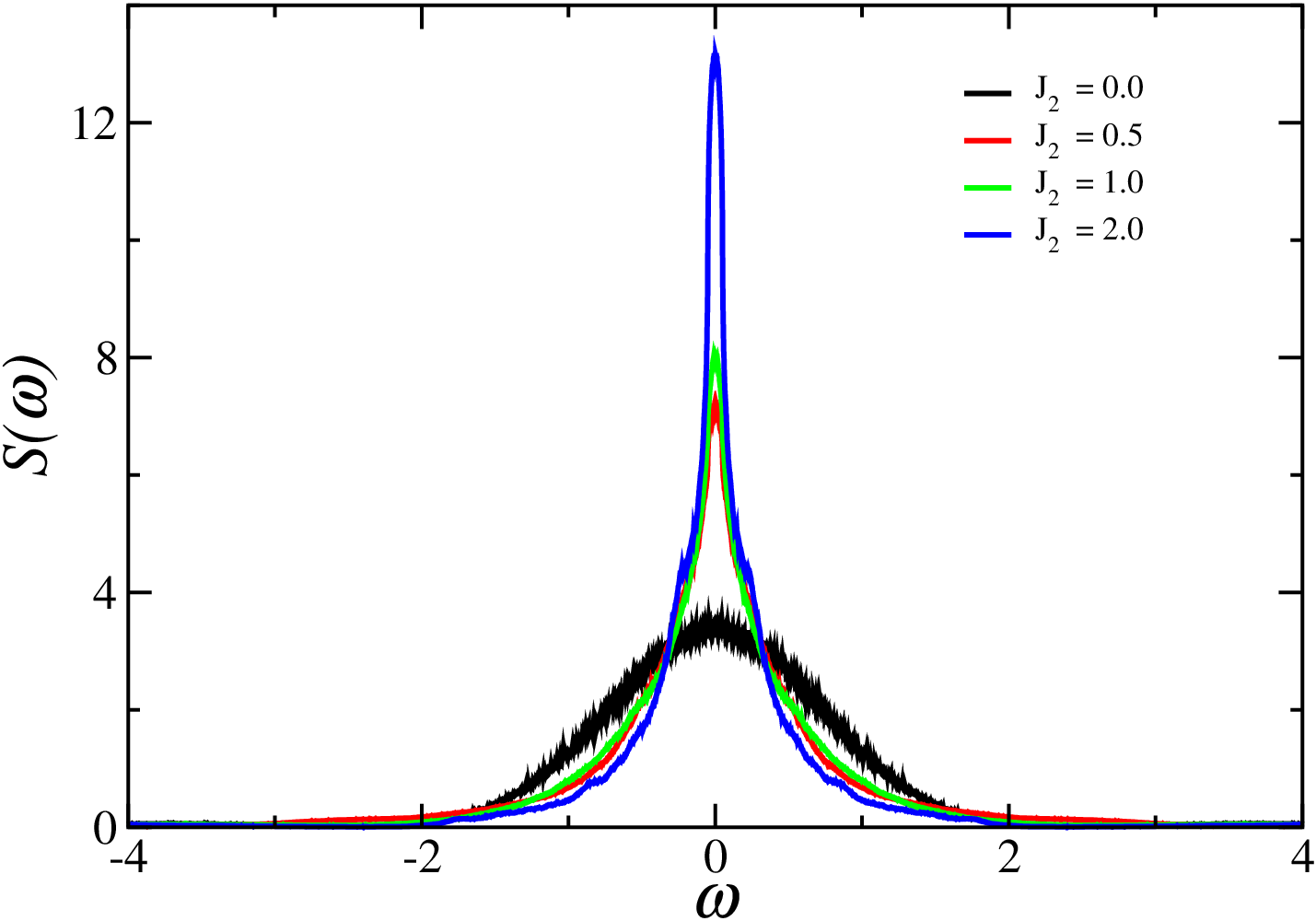}
\caption{Spectral density for $B = 0.5$ and several values of $J_2$.  All the curves were obtained
for chains with $L= 12$ spins.
\label{fig:fig4}
}
\end{figure}

The time correlation function for $B=1$ is depicted in
Fig.~\ref{fig:fig5} for several $J_2$.  The curves
shown are from $L=12$ and $L=13$.  Note that
for $J_2=0$ the calculation reproduces the known
Gaussian solution of the TI model. When $J_2=0.5$, oscillations are present in $C(t)$.  
For $J_2 \ge 1$ the curves decay at much slower rate.   A careful examination of
the figures shows oscillations of relatively small amplitudes.  
The spectral density $S(\omega)$
is shown in Fig.~\ref{fig:fig6}, where the calculations
were done with $L=13$.  For $J_2=0$ the Gaussian of the
TI model is reproduced.  We observe an enhancement of the
central model behavior as the values of $J_2$ are increased. 

Finally, consider the case where the transverse field  is larger ($B=2$) than
the Ising coupling.  The time correlation function
is shown in Fig.~\ref{fig:fig7} for several values
of $J_2$.  The curves were obtained from a chain
of size $L=13$.  For small values of $J_2$ the correlation function, $C(t)$, shows
oscillations typical of collective mode, such as
that found in the TI model ($J_2 = 0$).  As $J_2$ becomes
larger, the amplitude of the oscillations decreases.
For large enough $J_2$, the system displays
an enhancement of the central model.

Figure~\ref{fig:fig8}  depicts the corresponding spectral density $S(\omega)$ for
the values of $J_2$ used in the previous figure.
For $J_2=0$, the dynamics is dominated by the
two-peak structure characteristic of collective
mode. As $J_2$ increases, a  reduction of the intensity of the peaks 
of the collective mode is observed in tandem with a growth of the central peak.
For $J_2 \ge 2 $ the dynamics seems to be dominated entirely by the central mode.

The recurrants $\Delta_{\nu}$ of the method  of recurrence relations are calculated numerically
for $J_2=1$ and several values of $B$. First  the moments $\mu$ are obtained by using
Eq.~(\ref{eq:Sdirac}). Next, use the conversion formulas Eq.~(\ref{eq:Delta}).
Table~\ref{tab:tab1} shows some numerical results for the recurrants when $B=1.0$,
and $J_2=1.0$, obtained for $L= 11$, $12$, and $13$.  The rightmost column shows
the extrapolated value of $\Delta_{\nu}$  for $L= \infty$.  As can be seen, with relatively
small chain sizes ($L \le 13$), one can infer the thermodynamic value of the lower-order recurrants.
Higher order recurrants are still obtained, but with lesser accuracy.
 
The results for the thermodynamic estimates of the recurrants are shown in Fig.~\ref{fig:fig9}
for $B=1.0$ and various values for $J_2$. For $J_2=0$ the {\it linear} behavior that
leads to Gaussian behavior is recovered~\cite{Flo87}.
As $J_2$ increases, $\Delta_{\nu}$ increases at  higher rates on the average and 
becomes rather erratic, therefore, it is difficult to predict a trend based on their behavior.  
Still the results shown are already the thermodynamic values,
and it is very difficult to devise extrapolation schemes for the $\Delta$.
Notwithstanding, such an endeavor will not uncover any new physics in regard to the
dynamics of the transverse ANNNI model considered here.

\begin{figure}
\includegraphics[width=8.0cm, height= 6.0cm, angle=0]{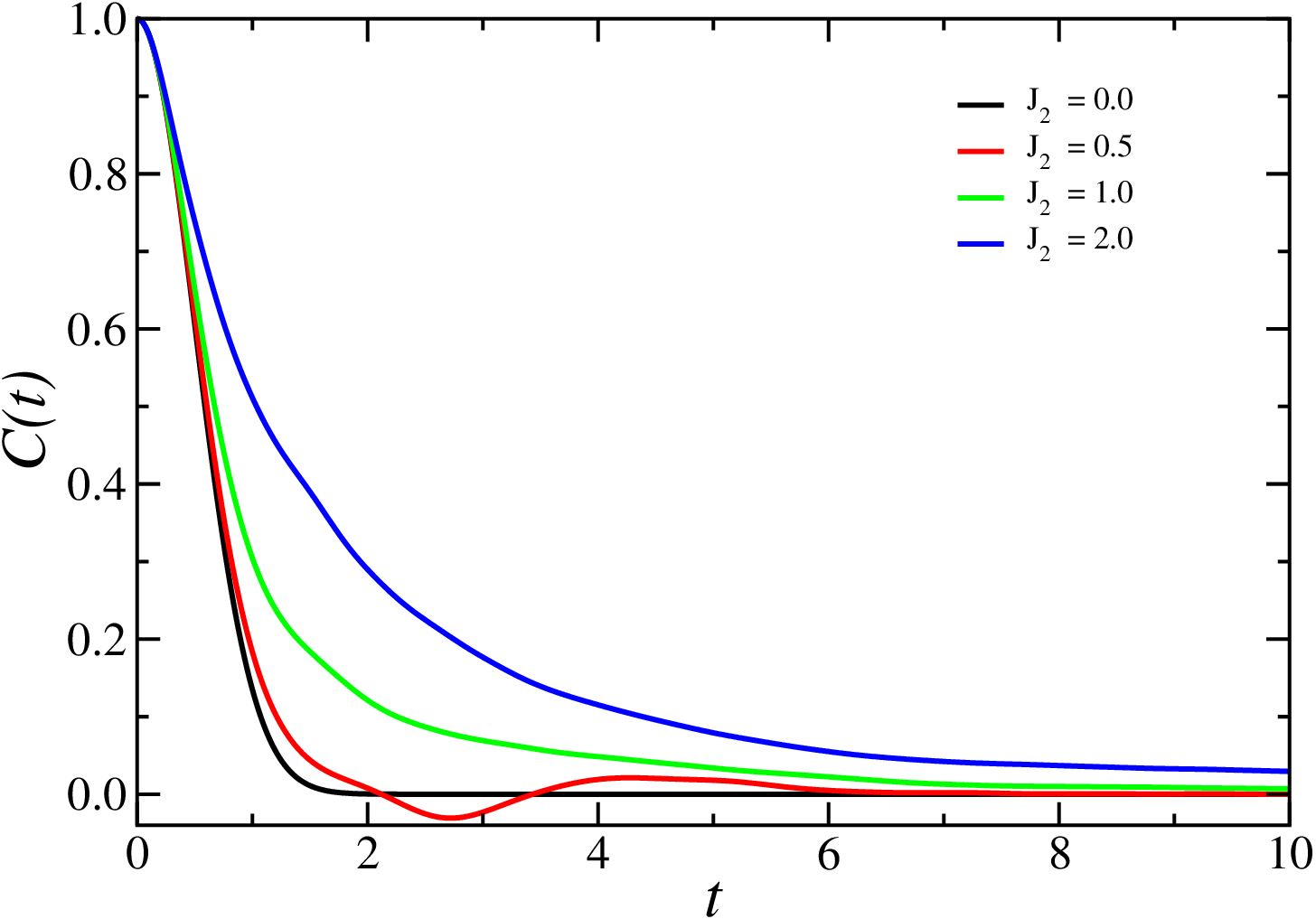}
\setlength{\belowcaptionskip}{30pt}
\caption{Time-dependent correlation function for $B = 1$ and several values
of $J_2$.  The curves were obtained with $L=12$ and $13$.
\label{fig:fig5}
}
\end{figure}

\begin{figure}
\includegraphics[width=8.0cm, height= 6.0cm, angle=0]{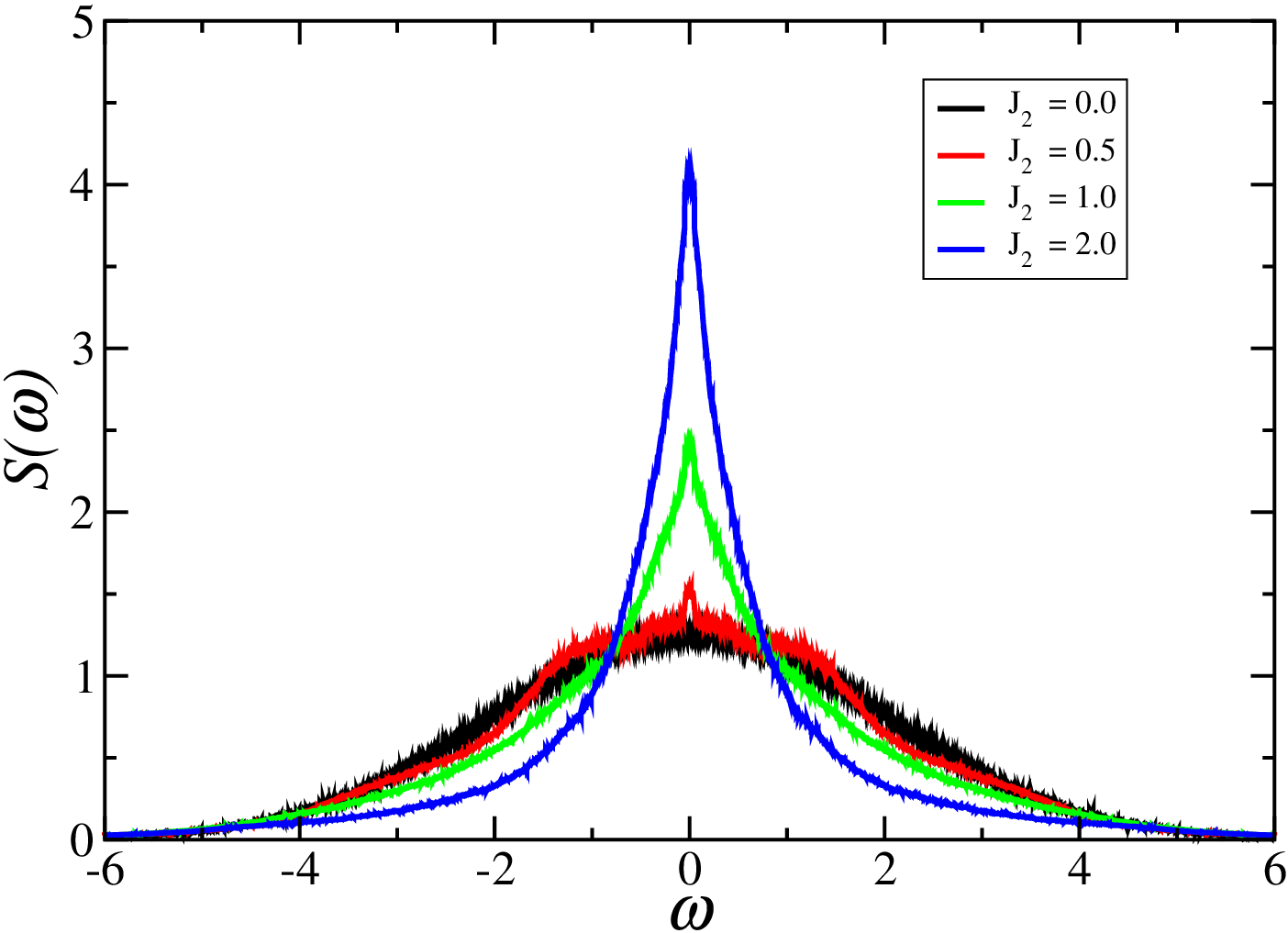}
\caption{Spectral density for  $B = 1$ and several values of $J_2$, obtained
for chains of length $L=13$.
\label{fig:fig6}
}
\end{figure}

\begin{figure}
\includegraphics[width=8.0cm, height= 6.0cm, angle=0]{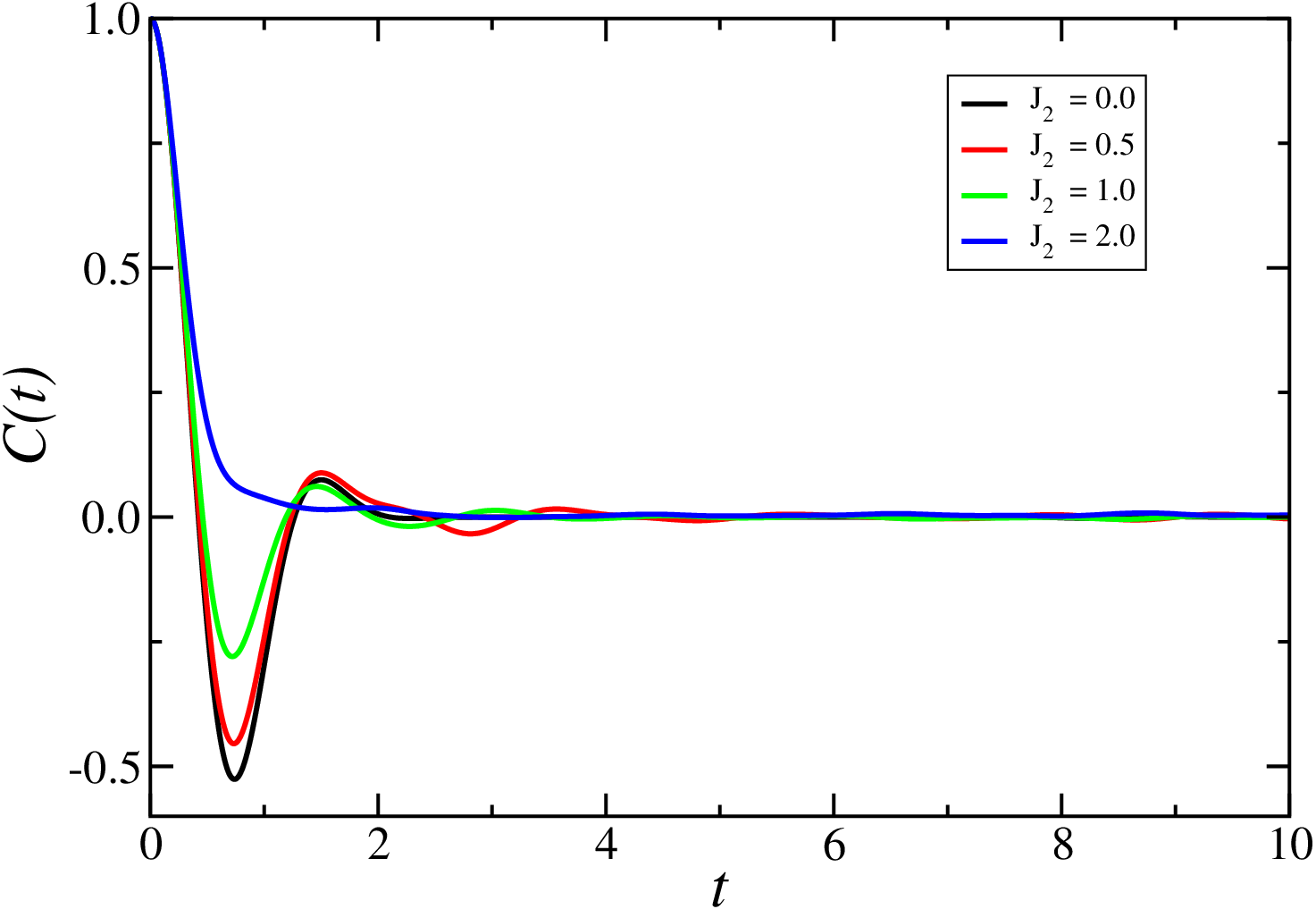}
\setlength{\belowcaptionskip}{20pt}
\caption{Time-dependent correlation function for $B = 2.0$ and various values of $J_2$.  
The chain size is $L=13$.
\label{fig:fig7}
}
\end{figure}

\begin{figure}
\includegraphics[width=8.0cm, height= 6.0cm, angle=0]{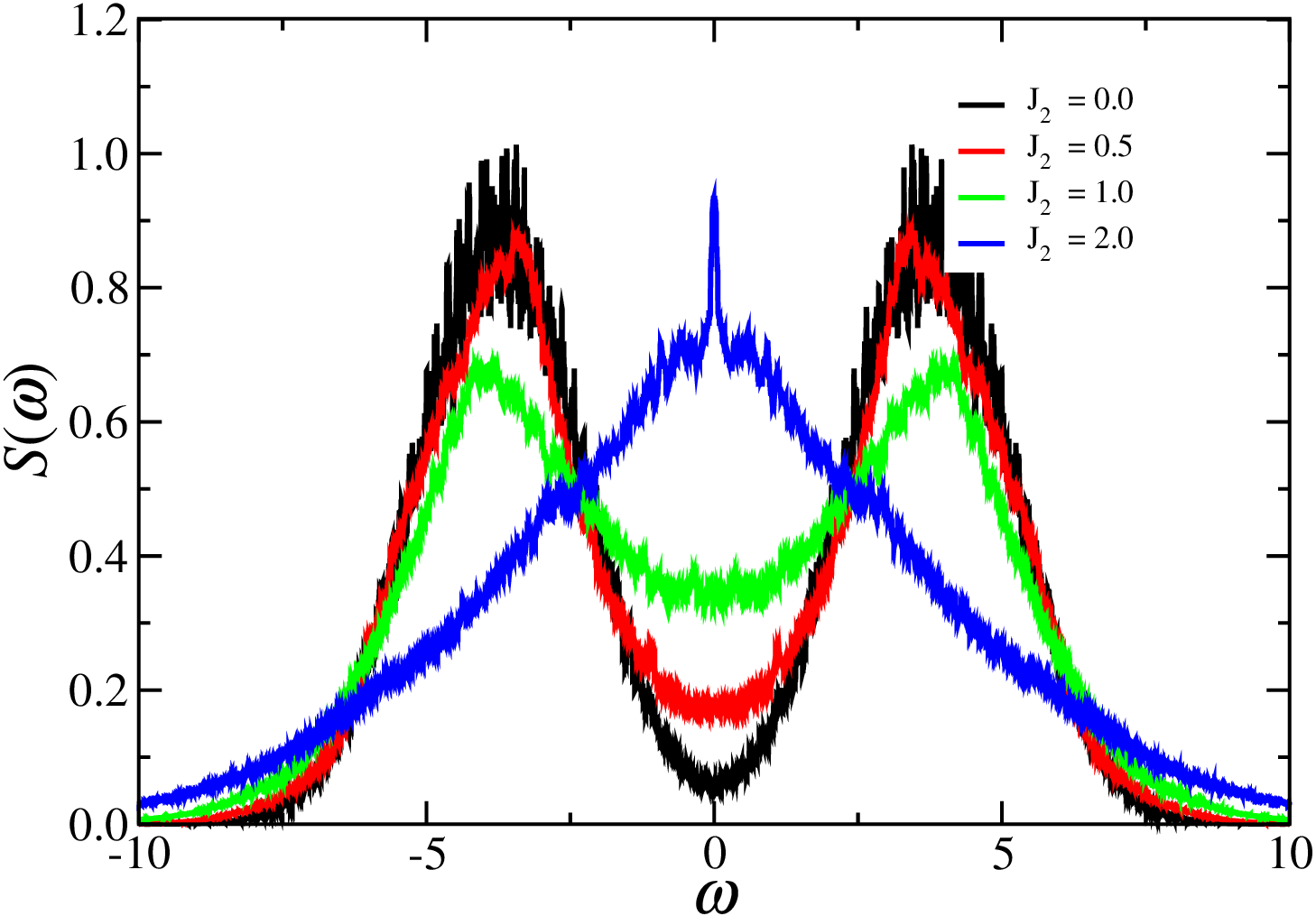}
\setlength{\belowcaptionskip}{20pt}
\caption{Spectral density for the case $B=2$ and several values of $J_2$.
The plots were obtained for $L=12$.
\label{fig:fig8}
}
\end{figure}

\begin{table}
 \begin{tabular}{lcccc}
$\Delta_{\nu} $  &   $  L=11  $ & $  L=12  $ & $  L=13  $ & $  L=\infty   $\\
 $\Delta_{1}   $ & $    4.00000   $ & $   4.00000   $ & $   4.00000   $ & $   4.00000   $ \\
 $\Delta_{2}   $ & $    16.0000   $ & $   16.0000   $ & $   16.0000   $ & $   16.0000   $ \\
 $\Delta_{3}   $ & $    28.0000   $ & $   28.0000   $ & $   28.0000   $ & $   28.0000   $ \\
 $\Delta_{4}   $ & $    41.1429   $ & $   41.1429   $ & $   41.1429   $ & $   41.1429   $ \\
 $\Delta_{5}   $ & $    51.3016   $ & $   51.3016   $ & $   51.3016   $ & $   51.3016   $ \\
 $\Delta_{6}   $ & $    73.0933   $ & $   73.0933   $ & $   73.0933   $ & $   73.0933   $ \\
 $\Delta_{7}   $ & $    78.5228   $ & $   78.5228   $ & $   78.5228   $ & $   78.5228   $ \\
 $\Delta_{8}   $ & $    92.4927   $ & $   92.4927   $ & $   92.4927   $ & $   92.4927   $ \\
 $\Delta_{9}   $ & $    110.406   $ & $   110.406   $ & $   110.406   $ & $   110.406   $ \\
 $\Delta_{10}   $ & $   127.334   $ & $   127.334   $ & $   127.334   $ & $   127.334   $ \\
 $\Delta_{11}   $ & $   151.014   $ & $   151.014   $ & $   151.014   $ & $   151.014   $ \\
 $\Delta_{12}   $ & $   168.388   $ & $   168.385   $ & $   168.385   $ & $   168.385   $ \\
 $\Delta_{13}   $ & $   191.746   $ & $   191.673   $ & $   191.672   $ & $   191.67    $ \\
 $\Delta_{14}   $ & $   216.807   $ & $   216.023   $ & $   215.961   $ & $   216.0     $ \\
 $\Delta_{15}   $ & $   233.220   $ & $   229.579   $ & $   229.217   $ & $   2.3\times 10^2       $ \\
 $\Delta_{16}   $ & $   269.065   $ & $   259.252   $ & $   258.141   $ & $   2.6\times 10^2       $ \\
 $\Delta_{17}   $ & $   298.445   $ & $   281.726   $ & $   278.903   $ & $   2.8\times 10^2       $ \\
    \end{tabular}
  \caption{Recurrants for the transverse ANNNI  model, with $ B=1.0$, $J_2=1.0$ and 
   several chain sizes. The rightmost column is the extrapolation for the thermodynamic
    limit ($L=\infty$).}
  \label{tab:tab1}
\end{table}

\begin{figure}
\includegraphics[width=8.0cm, height= 6.0cm, angle=0]{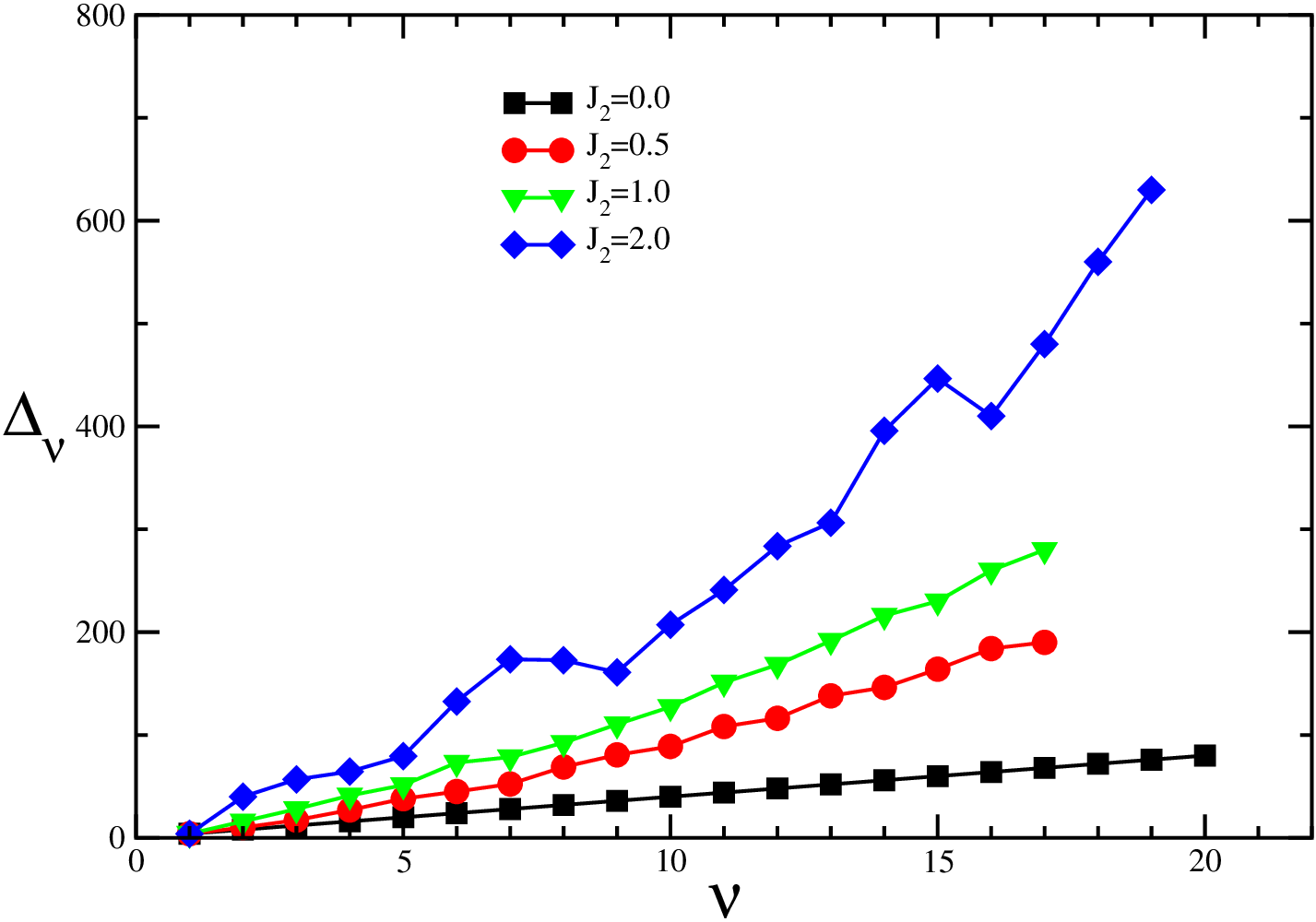}
\caption{Recurrants of the infinite transverse ANNNI model with  $B=1.0$ and several values of $J_2$.
\label{fig:fig9}
}
\end{figure}

\section{Summary and perspectives}

\noindent The dynamical correlation functions play a crucial role in the fluctuation-dissipation theorem
and in the linear response theory.  However, the calculation of those quantities is often
a very complicated problem in itself.  The method of recurrence relations is an exact
procedure that allows one to obtain of time correlation functions, spectral densities,
and dynamical structure factors.  We have shown the main features of the method and
the inherent difficulties one might encounter in an attempt to apply to a many-body problem.  
Another method that is showing great potential is exact diagonalization, a numerical method
which relies mostly on computer capabilities.  Nevertheless, the two methods can be
used together, one complementing the other, to achieve progress in the calculation
of dynamical correlation functions. 
 
\vskip 0.5cm

\noindent {\bf Acknowledgements}
\vskip 0.2cm
\noindent This work was supported by FAPERJ and PROPPI-UFF (Brazilian agencies).

\vskip 0.2cm
\section{Conflict of Interest Statement}
 
\vskip 0.2cm
\noindent The authors declare that the research was conducted in the absence of any commercial or financial relationships that could be construed as a potential conflict of interest.

\vskip 0.2cm
\section{Author Contributions}

\vskip 0.2cm
\noindent The authors J. F and O. F. A. B. declare that they planned and carried out
the elaboration of this work with equal contribution from each author.

\end{document}